\documentclass{svjour2}                    % onecolumn
\smartqed  % flush right qed marks, e.g. at end of proof
\begin{document}

\def \d {{\rm d}}
\def \im {{\rm i}} 
\def \boldk {\mbox{\boldmath$k$}} 
\def \boldl {\mbox{\boldmath$l$}} 
\def \boldm {\mbox{\boldmath$m$}} 
\def\op#1{\mathord{\rm #1}}

\title{On conformally flat and type N pure radiation metrics%\thanks{Grants or other notes
}


\author{Ji\v r\'i Podolsk\'y         \and
        Ond\v rej Prikryl %etc.
}


\institute{%F. Author \at
              Institute of Theoretical Physics, Faculty of Mathematics and Physics,\\
              Charles University in Prague, V Hole\v{s}ovi\v{c}k\'{a}ch 2, 180 00 Prague 8,  Czech Republic
              Tel.: +420-221912505, \email{podolsky`AT'mbox.troja.mff.cuni.cz} %
}

\date{Received: 22 August 2008 / Accepted: date}

\maketitle

\begin{abstract}
We study pure radiation spacetimes of algebraic types~O and~N with a possible cosmological constant. In particular, we present explicit transformations which put these metrics, that were recently re-derived by Edgar, Vickers and Machado Ramos, into a general Ozsv\'ath--Robinson--R\'ozga form. By putting all such metrics into the unified coordinate system we confirm that their derivation based on the GIF formalism is correct. We identify only few trivial differences. 

% 04.20.Jb  Exact solutions
% 04.30.Nk  Wave propagation and interactions
\end{abstract}

\section{Introduction}
\label{intro}
In recent papers \cite{EdgarVickers:1999,EdgarRamos:2005,EdgarRamos:2007a,EdgarRamos:2007b}, Edgar, Vickers and Machado Ramos systematically re-derived a general family of pure radiation spacetimes which are of algebraic types~N and~O. Using the advantages of the generalized invariant formalism (GIF) \cite{RamosVickers:1996}, which combines the features of standard Geroch--Held--Penrose (GHP) and null rotation invariant formalisms, they obtained a complete class of conformally flat and type~N solutions of the field equations, possibly admitting a cosmological constant $\Lambda$.

These metrics were presented in \cite{EdgarVickers:1999,EdgarRamos:2005,EdgarRamos:2007a,EdgarRamos:2007b} in the contravariant forms of $g^{ij}$ using various different coordinates. Our main aim here is to compare these (apparently distinct) metrics by putting all of them into a common coordinate system which is more suitable for physical and geometrical interpretation. This will also elucidate relations of these solutions to previous works, in particular those presented in \cite{OzsvathRobinsonRozga:1985,Siklos:1985,Podolsky:1998,BicakPodolsky:1999,EdgarLudwig:1997}. We will explicitly demonstrate that all such metrics can be conveniently written in the form
\begin{equation} 
 \d s^2=\frac{2}{P^2}\,\d\zeta\,\d\bar\zeta-2\,\frac{Q^2}{P^2}\,\d u\,\d v+\left(\kappa \,\frac{Q^2}{P^2}\,v^2-\frac{(Q^2)_{,u}}{P^2}\,v-\frac{Q}{P}\,H\right)\!\d  u^2\,,
 \label{metricORR}
\end{equation} 
where
\begin{eqnarray}
 P& =& 1+{\textstyle{1\over6}}\Lambda\zeta\bar\zeta \,, \label{coeffORR1}\\
 Q& =& (1-{\textstyle{1\over6}}\Lambda\zeta\bar\zeta )\,\alpha +\bar{\beta}\,\zeta+\beta\,\bar\zeta\,, \label{coeffORR2}
 \end{eqnarray}
 and
\begin{equation}
  \kappa  ={\textstyle\frac{1}{3}}\Lambda\, \alpha^2+2\,\beta\bar \beta\,,
\label{Kundtk}
 \end{equation}
where $\alpha(u)$, $\beta(u)$ are functions of $u$, and $H(\zeta,\bar\zeta,u)$ is a  function of ${\zeta, \bar\zeta, u}$. The general metric (\ref{metricORR})--(\ref{Kundtk}) was first presented by Ozsv\'ath, Robinson and R\'ozga in \cite{OzsvathRobinsonRozga:1985}. It represents all type N or conformally flat Kundt spacetimes with a cosmological constant which are either vacuum or contain pure radiation.  Indeed, with the  null tetrad  ${\,\boldk=\partial_v}$, ${\,\boldl=(P^2/Q^2)\,\partial_u+(P^4/2Q^4)F\,\partial_r}$, ${\,\boldm=P\partial_{\bar\zeta}\,}$ where ${F=g_{uu}}$, the only non-zero curvature tensor components are given by  
\begin{eqnarray}
  \Psi_4    &=&{\textstyle\frac{1}{2}}(P\,H)_{,\zeta\zeta}\, \frac{P^4}{Q^3}\,,\label{psiKundtN}\\
  \Phi_{22} &=&{\textstyle\frac{1}{2}}(P^2H_{,\zeta\bar\zeta} +{\textstyle\frac{1}{3}}\Lambda\, H )\frac{P^3}{Q^3} \,. \label{psiphiKundtN}
\end{eqnarray}

The complete family of {\em conformally flat\,} Kundt spacetimes with pure radiation are obtained by setting ${\,\Psi_4=0}$, in which case the function $H$ takes the particular form
\begin{equation}
H=\frac{{\cal A}(u)+\bar {\cal B}(u)\zeta+{\cal B}(u)\bar\zeta+{\cal 
C}(u)\zeta\bar\zeta}{1+{\textstyle{1\over6}}\Lambda\zeta\bar\zeta }\,,
\label{KundtConflat}
\end{equation}
where ${\cal A}(u)$, ${\cal B}(u)$ and ${\cal C}(u)$ are arbitrary functions of $u$, with ${\cal A}$ and ${\cal C}$ real. The pure radiation component is then
\begin{equation}
   \Phi_{22} = {\textstyle\frac{1}{2}}\left({\cal C} +{\textstyle{1\over6}}\Lambda\, {\cal A} 
\right)\frac{P^3}{Q^3} \,. \label{PhiKundtO}
\end{equation}

The spacetimes are {\em vacuum\,} when ${\,\Phi_{22}=0}$, i.e.  ${\,P^2H_{,\zeta\bar\zeta} +{\textstyle{1\over3}}\Lambda\, H=0\,}$. This equation has a general solution $ {\,H=(f_{,\zeta}+\bar f_{,\bar\zeta})-\frac{1}{3}\Lambda P^{-1}(\bar\zeta f + \zeta\bar f)\,}$,
where $f(\zeta,u)$ is an arbitrary function of $\zeta$ and $u$, holomorphic in $\zeta$. In all other cases, the spacetimes contain pure radiation  and are of type N or~O.

In particular, for conformally flat {\em and\,} vacuum spacetimes, ${\,{\cal C} =-{\textstyle\frac{1}{6}}\Lambda\, {\cal A}\,}$.  Such functions~$H$ of the form (\ref{KundtConflat}) correspond to  ${f= c_0(u)+c_1(u)\zeta+c_2(u)\zeta^2}$, where $c_i(u)$ are complex functions of $u$ related to ${\cal A}$ and ${\cal B}$. These solutions, with $f$ quadratic in $\zeta$,  are isometric to Minkowski (if~${\Lambda=0}$), de~Sitter (if~${\Lambda>0}$) and anti-de~Sitter spacetime (if~${\Lambda<0}$), see \cite{BicakPodolsky:1999}. For all other choices of $H$, the Kundt spacetimes (\ref{metricORR})--(\ref{Kundtk}) describe exact non-expanding pure radiation and/or gravitational waves.

\section{Type N spacetimes with pure radiation}
\label{sec:1}

The class of type N pure radiation solutions of Einstein's field equations with ${\Lambda=0}$ within the Kundt family was obtained in \cite{EdgarRamos:2005}. In the coordinates ${(t,n,a,b)}$ of equation (66) therein, the contravariant metric tensor reads\footnote{Here and in the following we are changing the signature to ${(-+++)}$.}
\begin{equation}
g^{ij} = 
\left( \begin{array}{cccc}
0    & 1/a  & 0   & 0 \\
1/a  & -L/a & n/a & 0 \\
0    & n/a  & 1   & 0 \\
0    & 0    & 0   & 1
\end{array} \right),
\label{Contravariant1}
\end{equation} 
where the arbitrary function ${L(t,\xi, \bar{\xi})}$ depends on three real coordinates ${t,a,b}$ via the complex variable ${\xi = a + \hbox{i}\,b}$. The inverse matrix to (\ref{Contravariant1}) is
\begin{equation}
g_{ij} =
\left( \begin{array}{cccc}
(n^2+aL) & a & -n & 0 \\
a      & 0 & 0  & 0 \\
-n     & 0 & 1  & 0 \\
0      & 0 & 0  & 1
\end{array} \right),
\label{Covariant1}
\end{equation}
so that the line element can be written as
\begin{equation}
\d s^2={\d a}^2+\d b^2-2n\, \d t\,\d a +2a\, \d t\,\d n +(n^2+aL)\,\d t^2.
\label{Metric1a}
\end{equation}
Now, performing the transformation and re-labelling
\begin{equation}
x=a, \qquad y=b, \qquad v=-\frac{n}{2a} , \qquad u=t,  
\label{sub1a}
\end{equation}
the metric becomes
\begin{equation}
\d s^2=\d x^2+\d y^2-4x^2\, \d u\,\d v +(4x^2v^2+x\,L)\,\d u^2,
\label{Metric1b}
\end{equation}
which is the standard explicit form of the Kundt type~N metrics, see e.g. \cite{PodolskyBelan:2004}. Introducing the complex spatial parameter ${\zeta=\frac{1}{\sqrt{2}}(x+\hbox{i}\,y)}$, we obtain
\begin{equation}
{\d s}^2=2 \, {\d \zeta}{\d \bar{\zeta}}-2(\zeta + \bar{\zeta})^2 \d u\,\d v + \left[2 (\zeta + \bar{\zeta})^2 v^2 - (\zeta + \bar{\zeta})\,H\right] \,{\d u}^2.
\label{Metric1c}
\end{equation}
This is the particular case ${\,P = 1}$, ${\,Q = \zeta + \bar{\zeta}}$, ${\,H = -  \frac{1}{\sqrt{2}}\,L\,}$
%\begin{equation}
%p = 1\,, \qquad q = \zeta + \bar{\zeta}\,, \qquad H = -  \textstyle{\frac{1}{\sqrt{2}}}\,L  \label{ozn1a}\,,
%\end{equation}
of the general metric  (\ref{metricORR}) of Kundt type~N spacetimes given in \cite{OzsvathRobinsonRozga:1985,BicakPodolsky:1999}, which corresponds to the choice of parameters ${\Lambda = 0}$, ${\alpha=0}$, ${\beta=1}$, i.e. ${\kappa=2}$.

 \section{Type~O spacetimes with pure radiation}

Next, we will examine the family of solutions obtained in~\cite{EdgarVickers:1999}. In the coordinates ${(t,n,a,b)}$, these conformally flat spacetimes are given by 
\begin{equation}
\it{g^{ij}} = 
\left( \begin{array}{cccc}
0    & 1/a                 & 0   & 0 \\
1/a\,  & \,(2S-2Ma-a^2-b^2)/a\, & \,n/a\, &\, E/a \\
0    & n/a                 & 1   & 0 \\
0    & E/a                 & 0   & 1
\end{array} \right),
\label{Contravariant0}
\end{equation} 
where  ${E, M, S}$ are arbitrary functions of $t$ (see equation (68) in ~\cite{EdgarVickers:1999}). The corresponding covariant metric tensor components are obtained by inverting this matrix, i.e. 
\begin{equation}
\it{g_{ij}} =
\left( \begin{array}{cccc}
(n^2+E^2+a L)\, &\, a\, &\, -n\, &\, -E \\
a                           & 0 & 0  & 0 \\
-n                          & 0 & 1  & 0 \\
-E                          & 0 & 0  & 1
\end{array} \right),
\label{Covariant0}
\end{equation}
where 
\begin{equation}
 L \equiv a^2+b^2+2Ma-2S.
\label{ozn0a}
\end{equation}
The metric thus takes the explicit form

\begin{equation}
{\d s}^2={\d a}^2+{\d b}^2-2n\, \d t\,\d a -2E\, \d t\,\d b +2a\, \d t\,\d n +(n^2+E^2+aL)\,{\d t}^2,
\label{Metric0a}
\end{equation}
which, for ${E=0}$, obviously reduces to (\ref{Metric1a}). With the transformation and re-labelling
\begin{equation}
x=a, \qquad  y=b-\int\! E \,\d t, \qquad v=-\frac{n}{2a} , \qquad u=t,  
\label{sub0a}
\end{equation}
the metric becomes
\begin{equation}
\d s^2=\d x^2+\d y^2-4x^2\, \d u\,\d v +(4x^2v^2+x\,L)\,\d u^2,
\label{Metric0b}
\end{equation}
where the function $L$ is
\begin{equation}
 L = {\textstyle x^2+y^2+2M\,x+\left(2\int \!E \, \d u\right)y -2S+ \left(\int E \, \d u\right)^2}.
\label{ozn0b}
\end{equation}
This is again the standard Kundt metric (\ref{Metric1b}), but the function $L$ now has a special form (\ref{ozn0b}). It is fully consistent with the assumed conformal flatness which requires $L$ to be at most quadratic in the spatial coordinates $x$ and $y$, with the coefficients being arbitrary functions of $u$. Introducing the complex coordinate ${\zeta=\frac{1}{\sqrt{2}}(x+\hbox{i}\,y)}$, we obtain the metric (\ref{Metric1c}) in which $L$ is 
\begin{equation}
 L(\zeta,\overline{\zeta},u) = 2\,\zeta\bar{\zeta} +f(u)\,\zeta+g(u)\,\bar{\zeta}+h(u),
\label{ozn0c}
\end{equation}
where  $f,g,h$ are any functions of~$u$. This exactly corresponds to the function $H$ given by expression (\ref{KundtConflat}) in the case when ${\Lambda=0}$ and ${{\cal C}(u)}$ is constant.

 \section{Type~O pure radiation spacetimes with a negative cosmological constant}
 \label{sectypeonegat}

Now, we will analyse the solutions described by equation (103) in \cite{EdgarRamos:2007a} which represent a class of conformally flat  pure radiation metrics with ${\Lambda<0}$ such that ${\tau\bar{\tau}+\frac{1}{6}\Lambda=0}$, where $\tau$ is the corresponding spin coefficient. The contravariant metric tensor components in coordinates ${(r,n,m,b)}$ are
\begin{equation}
g^{ij} = m^2 \,
\left( \begin{array}{cccc}
0    & -1        & 0           & 0 \\
-1   & -V        & -m\,\nu_4(r)  & -b\,\nu_4(r) \\
0    & -m\,\nu_4(r) & 2\lambda^2 & 0 \\
0    & -b\,\nu_4(r) & 0          & 2\lambda^2
\end{array} \right),
\label{Contravariant2}
\end{equation} 
where 
\begin{equation}
 V = 3n\,\nu_4(r) - \nu_5(r)(b^2+m^2) - \nu_6(r)\,b - \frac {m}{\lambda^2} + \nu_3(r)\,,
\label{ozn2a}
\end{equation}
in which  ${\nu_3(r),\, \nu_4(r),\, \nu_5(r),\, \nu_6(r)}$ are arbitrary functions of~$r$. The inverse matrix is
\begin{equation}
g_{ij} = \frac {1}{2\lambda^2m^2} \,
\left( \begin{array}{cccc}
-\tilde V    & -2\lambda^2 & \,-{m\,\nu_4(r)}\, & \,-{b\,\nu_4(r)}\, \\
-2\lambda^2  & 0   & 0   & 0 \\
-{m\,\nu_4(r)} & 0   & 1   & 0 \\
-{b\,\nu_4(r)} & 0   & 0   & 1
\end{array} \right),
\label{Covariant2}
\end{equation}
with 
\begin{equation}
 \tilde V \equiv -2\lambda^2 V - \nu_4^2(r)(m^2+b^2)\,,
\label{ozn2b}
\end{equation}
so that the metric reads
\begin{eqnarray} 
&&\d s^2=\frac {1}{2\lambda^2m^2}\, \Big({\d m}^2+{\d b}^2 -2m\nu_4(r)\, \d r\,\d m -2 b\,\nu_4(r)\, \d r\,\d b \nonumber\\
&&\hspace{60mm}- 4\lambda^2\, \d r\,\d n  - \tilde V\,{\d r}^2 \Big).
\label{Metric2a}
\end{eqnarray}
Applying the transformation
\begin{equation}
x = m\,R^2(r)\,, \qquad
y = b\,R^2(r)\,, \qquad
v = n\,R^3(r)\,,  \label{sub2a}
\end{equation}
where
\begin{equation}
R(r) = \exp{\Big(-\frac{1}{2}\int\! \nu_4(r)\, \d r\Big)}, \label{sub2aa}
\end{equation}
we obtain
\begin{equation}
\d s^2=\frac {1}{2\lambda^2x^2}\, \left( \d x^2 + \d y^2 - 4\lambda^2 R(r)  \, \d r\,\d v + 2\lambda^2 \tilde H \,{\d r}^2 \right),
\label{Metric2b}
\end{equation}
in which $\tilde H$ has the form
\begin{equation}
 \tilde H \equiv - \nu_5(r)(x^2+y^2) - \nu_6(r) R^2(r)\,y  - \frac {R^2(r)}{\lambda^2}\,x  + \nu_3(r)R^4(r)\,.
\label{ozn2css}
\end{equation}
Finally, the transformation from $r$ to the new coordinate $u$,
\begin{equation}
 u = 2 \lambda^2 \int\! R(r)  \, \d r\,,
\label{sub2b}
\end{equation}
puts the above metric to
\begin{equation}
{\d s}^2=\frac {1}{2 \lambda^2 x^2}\, \left( \d x^2 + \d y^2 - 2 \, \d u\,\d v + H \,{\d u}^2 \right),
\label{Metric2c}
\end{equation}
where the function $H$ is obtained from $\tilde H$ by substitution for $r$,
\begin{equation}
  H(x,y,u)  = A(u)(x^2+y^2) + B(u)\,x + C(u)\,y  + D(u)\,,
\label{ozn2c}
\end{equation}
in which $A,B,C,D$ are arbitrary\footnote{Notice that during the derivation we have obtained ${B(u)=-1/(2\lambda^4)}$. However, an arbitrary function $B(u)$ can be set to any constant value by the coordinate transformation ${x=e^fx'}$, ${y=e^fy'}$, ${u=\int e^{2f}\d u' }$, ${v=v'+\frac{1}{2}\dot f( x'^2+y'^2)}$, using a suitable function $f(u')$, see \cite{Siklos:1985}.} functions of $u$. With the identification 
\begin{equation}
 \lambda^2 \equiv - \frac {\Lambda}{6},
\label{ozn2d}
\end{equation}
where ${\Lambda<0}$ is a negative cosmological constant, this is exactly the conformally flat subfamily of the Siklos solutions presented for the first time in \cite{Siklos:1985}. Using the transformation ${\zeta=-\sqrt{-{6\over\Lambda}}\,(x+{1\over2}+\im\, y)/(x-{1\over2}+\im\,y)}$, ${\,v={12\over\Lambda}\,r}$, see \cite{Podolsky:1998}, the metric (\ref{Metric2c}) is put into the Ozsv\'ath--Robinson--R\'ozga form (\ref{metricORR})--(\ref{KundtConflat}) with ${\alpha=1}$, ${\beta=\sqrt{-{1\over6}\Lambda}}$, i.e. ${Q=\left(1+\sqrt{-{1\over6}\Lambda}\,\zeta\right)\!\left(1+\sqrt{-{1\over6}\Lambda}\,\bar\zeta\right)}$ and ${\,\kappa=0}$.

\section{Other type~O pure radiation spacetimes with a cosmological constant}
 \label{sectypeogener}
 
In recent work \cite{EdgarRamos:2007b}, Edgar and Machado Ramos extended their investigations to a complete family of conformally flat  pure radiation metrics with ${\Lambda\not=0}$ for which ${\tau\bar{\tau}+\frac{1}{6}\Lambda\not=0}$. The corresponding contravariant metric tensor components given by equation (86) therein, using the coordinates ${(\tilde{t},c,a,\tilde{x})}$, are
\begin{equation}
\it{g^{ij}} = \left( \begin{array}{cccc}
0                                   & -\frac {k^{1/4}}{a(\frac {3}{2}-k)}                    & 0                                                & 0 \\
&&&\\
-\frac {k^{1/4}}{a(\frac {3}{2}-k)} \quad& \frac {8}{ak^{1/2}(\frac {3}{2} - k)^2}Z    \quad & \frac {2k(\frac{5}{36}\Lambda^2a^4+1)c}{a(\frac {3}{2}-k)} \quad& -\frac {k^{1/4}\gamma_3(\tilde{t})}{a(\frac {3}{2}-k)} \\
&&&\\
0                                   & \frac {2k(\frac{5}{36}\Lambda^2a^4+1)c}{a(\frac {3}{2}-k)}        & 4k^2                                             & 0 \\
&&&\\
0                                   & -\frac {k^{1/4}\gamma_3(\tilde{t})}{a(\frac {3}{2}-k)} & 0                                                & 4k
\end{array} \right),
\label{Contravariant3}
\end{equation} 
where $\gamma_3(\tilde{t})$ is an arbitrary function of $\tilde{t}$, and $k$ is given by\footnote{We have re-labeled the cosmological parameter $\Lambda$ used in \cite{EdgarRamos:2007b} to ${\textstyle \frac{1}{6}\Lambda}$, where $\Lambda$ is now the standard cosmological constant. See also relation (\ref{ozn2d}).}
\begin{equation}
 k = {\textstyle \frac{1}{6}\Lambda\, a^2 + \frac {1}{2}}\,.
\label{ozn3a}
\end{equation}
The function $Z$ may have three distinct forms, namely
\begin{eqnarray}
\hbox{(i) for ${\Lambda>0}$:} \ && Z = \gamma_1(\tilde{t})\,\cos\Big(\sqrt{{\textstyle \frac{1}{6}}\Lambda}\,\tilde{x}\Big) + 2\sqrt{2}\,a\,\gamma_2(\tilde{t}) +z(a,c)
\,,\label{ozn3b}\\
\hbox{(ii) for ${\Lambda<0}$:} \ && Z = \gamma_1(\tilde{t})\,\cosh\Big(\sqrt{{\textstyle -\frac{1}{6}}\Lambda}\,\tilde{x}\Big) + 2\sqrt{2}\,a\,\gamma_2(\tilde{t}) +z(a,c)
\,,\label{ozn3c}\\
\hbox{(iii) for ${\Lambda<0}$:} \ &&Z = \!\pm\big(\!{{\textstyle \frac{-1}{24}}\Lambda}\big)^{-\frac{1}{2}}\!\exp\Big(\!\mp{\!\sqrt{{\textstyle -\frac{1}{6}}\Lambda}\,\tilde{x}}\Big) + 2\sqrt{2}\,a\,\gamma_2(\tilde{t}) + z(a,c)
\,.\qquad\label{ozn3d}
\end{eqnarray}
Here $\gamma_1(\tilde{t})$ and $\gamma_2(\tilde{t})$ are arbitrary functions, and
\begin{equation}
z(a,c) \equiv - \frac {12\,k^{1/2}}{\Lambda} + {\textstyle \frac {1}{288}\Lambda^2k^{1/2}a^3c^2(\frac{25}{36}\Lambda^2a^4 - \frac{1}{3}\Lambda a^2 + 13)}\,. 
\label{subst}
\end{equation}
The corresponding inverse matrix to (\ref{Contravariant3}) is
\begin{equation}
\it{g_{ij}} = \left( \begin{array}{cccc}
-\frac{8a}{k}Z+\frac{\gamma_3^2(\tilde{t})}{4k}+\frac{(\frac{5}{36}\Lambda^2a^4+1)^2c^2}{k^{1/2}} \quad& \frac {a(k-\frac{3}{2})}{k^{1/4}} \quad& \frac {(\frac{5}{36}\Lambda^2a^4+1)c}{2k^{5/4}} \quad& -\frac {\gamma_3(\tilde{t})}{4k} \\
&&&\\
 \frac {a(k-\frac{3}{2})}{k^{1/4}}      & 0                 & 0                    & 0 \\
&&&\\
\frac {(\frac{5}{36}\Lambda^2a^4+1)c}{2k^{5/4}}    & 0      & \frac {1}{4k^2}      & 0 \\
&&&\\
-\frac {\gamma_3(\tilde{t})}{4k}        & 0                 & 0                    & \frac {1}{4k}
\end{array} \right),
\label{Covariant3}
\end{equation}
so that the metric can be expressed as
\begin{eqnarray} 
&&{\d s}^2=\frac {1}{4k^2}\,{\d a}^2 + \frac{1}{4k}{\d \tilde{x}}^2  + \frac{(\frac{5}{36}\Lambda^2a^4+1)c}{k^{5/4}}\, \d \tilde{t}\,\d a - \frac{\gamma_3(\tilde{t})}{2k}\,\d \tilde{t}\,\d \tilde{x} \label{Metric3a}\\
&& \hskip 12mm + \frac{a(2k-3)}{k^{1/4}}\,\d \tilde{t}\,\d c + \left(-\frac{8a}{k}\,Z+\frac{\gamma_3^2(\tilde{t})}{4k}+\frac{(\frac{5}{36}\Lambda^2a^4+1)^2c^2}{k^{1/2}}  \right){\d \tilde{t}}^2\,.
 \nonumber
\end{eqnarray}
It is possible to apply the transformation
\begin{eqnarray}
&&   x = \frac {1}{\sqrt{2}} \left( \tilde{x} - \int\!\gamma_3(\tilde{t})\, \d \tilde{t} \right), 
\label{sub3a}\\
&& {v} =-\frac{1}{2\kappa}\, k^{3/4}\,\frac {\frac{1}{6}\Lambda\, a^2 - 1}{a} \, c\,, 
\label{sub3b}\\
&& {u} = 4\, \tilde{t}\,, 
\label{sub3c}
\end{eqnarray}
where $\kappa$ is a suitable non-vanishing constant. This puts the metric (\ref{Metric3a}) to a more compact form
\begin{equation} 
{\d s}^2=\frac {1}{4k^2}\,{\d a}^2 + \frac{1}{2k}{\d x}^2 - 2\, \frac{\kappa a^2}{2k\,\,}\, \d {u}\,\d {v} + \left(\!\kappa\,\frac{\kappa a^2}{2k\,\,}\, {v}^2 -\frac{a}{2k} \tilde{H}  \right){\d {u}}^2,
\label{Metric3d}
\end{equation}
in which the function $\tilde{H}(a,x,{u})$ reads
\begin{eqnarray}
\hbox{(i)}  && \tilde{H} = \gamma_1({u})\cos\Big(\!\sqrt{{\textstyle \frac{1}{6}}\Lambda}\,( \sqrt{2}\,x + {\textstyle \int\!{\gamma_3}({u}) \d {u} })\Big) + 2\sqrt{2}\,a\,\gamma_2({u}) - \frac {12k^{1/2}}{\Lambda}
\,,\nonumber\\
\hbox{(ii)}  && \tilde{H} = \gamma_1({u})\cosh\Big(\sqrt{{\textstyle -\frac{1}{6}}\Lambda}\,( \sqrt{2}\,x + {\textstyle \int\!{\gamma_3}({u}) \d {u} })\Big) + 2\sqrt{2}\,a\,\gamma_2({u}) - \frac {12k^{1/2}}{\Lambda}
\,,\nonumber\\
\hbox{(iii)} &&\tilde{H} = \!\pm\big(\!{{\textstyle \frac{-1}{24}}\Lambda}\big)^{-\frac{1}{2}}\!\exp\!\Big(\!\!\mp{\!\sqrt{{\textstyle -\frac{1}{3}}\Lambda}\,}\,x \Big)\exp\Big({\textstyle\mp\sqrt{-\frac{1}{6}\Lambda}
\int\!{\gamma_3}({u}) \d {u} }\Big) \nonumber\\
&&\hskip60mm+ 2\sqrt{2}a\gamma_2({u}) - \frac {12k^{1/2}}{\Lambda}
\,. \label{specialcaseiii}
\end{eqnarray}

Now, by comparing (\ref{Metric3d}) with (\ref{metricORR}) we immediately conclude that the two metrics are identical provided
\begin{eqnarray}
\frac{\kappa a^2}{2k\,\,} & = & \frac{Q^2}{P^2}\,, 
\label{trsub1}\\
\frac {1}{4k^2}\,{\d a}^2 + \frac{1}{2k}{\d x}^2 & = & \frac{2}{P^2}\,\d\zeta\,\d\bar\zeta\,, 
\label{trsub2}\\
 \frac{a}{2k} \tilde{H}  & = & \frac{Q}{P}\,H\,,
\label{trsub3}
\end{eqnarray}
with constant $\alpha$, $\beta$. In view of (\ref{ozn3a}), the condition (\ref{trsub1}) leads to the relation 
\begin{equation}
a=\frac{Q}{\sqrt{\kappa P^2-\frac{1}{3}\Lambda Q^2}}\,,
\label{agenerally}
\end{equation}
where $P$ and $Q$ are given by (\ref{coeffORR1}), (\ref{coeffORR2}). The condition (\ref{trsub2}) can then be used to find the relation between the real coordinate $x$ and the complex coordinate~$\zeta$. In fact, it can thus be shown that both relations (\ref{trsub1}) and (\ref{trsub2}) are satisfied if  
\begin{eqnarray}
a & = & \frac{1}{\sqrt{2}}\,\frac{(1-{\textstyle{1\over6}}\Lambda\zeta\bar\zeta )\,\alpha +\bar{\beta}\,\zeta +\beta\,\bar\zeta} {\sqrt{({\textstyle{1\over6}}\Lambda\bar\beta\zeta^2+ {\textstyle{1\over3}}\Lambda\alpha\zeta-\beta) 
 ({\textstyle{1\over6}}\Lambda\beta\bar\zeta^2+ {\textstyle{1\over3}}\Lambda\alpha\bar\zeta-\bar\beta)}}\,, 
\label{transub1}\\
x & = & \im \frac{\sqrt{\kappa}}{2}\,\big(F(\zeta)-\bar F(\bar\zeta)\big)\,,\quad \hbox{where}\ \>
F(\zeta)=\int\!\!\frac{\d\zeta}{{\textstyle{1\over6}}\Lambda\bar\beta\zeta^2+ {\textstyle{1\over3}}\Lambda\alpha\zeta-\beta}\,, 
\label{transub2}
\end{eqnarray}
with ${\, \kappa  ={\textstyle\frac{1}{3}}\Lambda\, \alpha^2+2\,\beta\bar \beta\not=0\,}$, cf. (\ref{Kundtk}). Of course, the function $F$ can be  integrated as
\begin{eqnarray}
F(\zeta) & = & \frac{3}{\Lambda \,\alpha}\,\ln\zeta \hskip11.0pc \hbox{for}\quad \beta=0\,, 
\label{intransub1}\\
F(\zeta)  & = &-2\sqrt{\frac{3}{\Lambda\kappa}} \,\,\hbox{arctanh}\,\Big(\sqrt{\frac{\Lambda}{3\kappa}}\,(\bar \beta\,\zeta+\alpha)\Big) \hskip1.3pc \hbox{for}\quad \beta\not=0 \,. 
\label{inttransub2}
\end{eqnarray}
The transformation (\ref{transub1}), (\ref{transub2}) puts the line element (\ref{Metric3d}) explicitly to the Ozsv\'ath--Robinson--R\'ozga metric form (\ref{metricORR}), namely
\begin{equation} 
 \d s^2=\frac{2}{P^2}\,\d\zeta\,\d\bar\zeta-2\,\frac{Q^2}{P^2}\,\d u\,\d v+\left(\kappa \,\frac{Q^2}{P^2}\,v^2-\frac{Q}{P}\,H\right)\!\d  u^2\,,
 \label{metricORRspec}
\end{equation} 
where $ {\,P = 1+{\textstyle{1\over6}}\Lambda\zeta\bar\zeta \,}$ and ${\, Q = (1-{\textstyle{1\over6}}\Lambda\zeta\bar\zeta )\,\alpha +\bar{\beta}\,\zeta+\beta\,\bar\zeta\,}$, with constant $\alpha, \beta$.

Moreover, using (\ref{trsub3}), the function $H$ can be expressed as 
\begin{equation}
H=\frac{\sqrt{2}}{\kappa \,P}\,{\sqrt{({\textstyle{1\over6}}\Lambda\bar\beta\zeta^2+ {\textstyle{1\over3}}\Lambda\alpha\zeta-\beta) 
 ({\textstyle{1\over6}}\Lambda\beta\bar\zeta^2+ {\textstyle{1\over3}}\Lambda\alpha\bar\zeta-\bar\beta)}}\,\tilde H\,,
\label{KundtConflattransf}
\end{equation}
In view of the above three possible forms (i)--(iii) of  $\tilde H$, and considering relations (\ref{agenerally})--(\ref{inttransub2}), it can be argued that the function $H$ takes the form (\ref{KundtConflat}) for the most general conformally flat solution of the Ozsv\'ath--Robinson--R\'ozga family of spacetimes.

As shown in \cite{OzsvathRobinsonRozga:1985,BicakPodolsky:1999}, there are various geometrically distinct subclasses of solutions represented by the metric (\ref{metricORRspec}). These subclasses can be distinguished by the cosmological constant $\Lambda$ and sign of the function~$\kappa$ defined in (\ref{Kundtk}). Each is represented by the corresponding canonical choice of the parameters $\alpha$ and~$\beta$. The case ${\kappa=0}$, which corresponds to the Siklos spacetimes with ${\tau\bar{\tau}+\frac{1}{6}\Lambda=0}$ and ${\Lambda<0}$, was described in previous section~\ref{sectypeonegat}. 
The solutions in the present section~\ref{sectypeogener} are characterised by ${\tau\bar{\tau}+\frac{1}{6}\Lambda\not=0}$, and correspond to the cases ${\kappa\not=0}$, ${\Lambda\not=0}$.

For ${\Lambda>0}$, there is only the subclass ${\kappa >0}$ of possible solutions, see (\ref{Kundtk}). Its canonical representation is given by ${\alpha=0}$, ${\beta=1}$, so that $ {\,P = 1+{\textstyle{1\over6}}\Lambda\zeta\bar\zeta \,}$, ${Q = \zeta + \bar\zeta\,}$, and ${\kappa = 2}$. In this case the transformation (\ref{transub1}), (\ref{transub2}) simplifies to
\begin{eqnarray}
  a &=&   \frac{\frac{1}{\sqrt{2}}(\zeta + \bar\zeta)}{\sqrt{(1-{\textstyle{1\over6}}\Lambda\zeta^2)(1-{\textstyle{1\over6}}\Lambda\bar\zeta^2)}}, \label{acase1}\\
  x &=& - \im\sqrt{\frac{3}{\Lambda}} \left(\hbox{arctanh} \Big(\!\sqrt{{\textstyle{1\over6}}\Lambda} \,\zeta\Big) 
   - \hbox{arctanh} \Big(\!\sqrt{{\textstyle{1\over6}}\Lambda}\, \bar\zeta\Big) \right), \label{xcase1}\\
    &=&  - \frac{\im}{2}\sqrt{\frac{3}{\Lambda}} \,\,\ln\frac{\left(1+\sqrt{\frac{1}{6}\Lambda}\,\zeta\right)\!\!\left(1-\sqrt{\frac{1}{6}\Lambda}\,\bar\zeta\right)}
{\left(1+\sqrt{\frac{1}{6}\Lambda}\,\bar\zeta\right)\!\!\left(1-\sqrt{\frac{1}{6}\Lambda}\,\zeta\right)}  \,. \nonumber
\end{eqnarray}
A straightforward calculation now shows that the function $H$ given by (\ref{KundtConflattransf}) takes exactly the form (\ref{KundtConflat}) where
\begin{eqnarray}
 && {\cal A}(u) = \frac{1}{\sqrt2}  \left(\gamma_1 (u)\,\cos\Big(\sqrt{{\textstyle \frac{1}{6}}\Lambda}\,{\textstyle \int\!{\gamma_3}({u}) \,\d {u} }\Big) - \sqrt{2}\,\frac{6}{\Lambda} \right), \label{Cverse2a}\\
 && {\cal B}(u) = \frac{1}{\sqrt2} \left( 2 \gamma_2 (u) - \im\sqrt{{\textstyle \frac{1}{6}}\Lambda}\,\gamma_1 (u)\,\sin\Big(\sqrt{{\textstyle \frac{1}{6}}\Lambda}\,{\textstyle \int\!{\gamma_3}({u}) \,\d {u} }\Big) \right), \label{Cverse2b}\\
 && {\cal C}(u) = -{\textstyle \frac{1}{6}}\Lambda\, {\cal A}(u) - 2\,, \label{Cverse2c}
\end{eqnarray}
which follows from the case (i) of  $\tilde H$.

If ${\Lambda<0}$, there are two subclasses of possible solutions for ${\kappa\not=0}$. When ${\kappa >0}$, the canonical representation is again given by ${\alpha=0}$, ${\beta=1}$, and the transformation has the form (\ref{acase1}), (\ref{xcase1}), with the replacement ${\,\Lambda\to-\Lambda\,}$  and ${\,\hbox{arctanh}\to\arctan\,}$ in~(\ref{xcase1}). Similarly, the trigonometric functions in (\ref{Cverse2a})--(\ref{Cverse2c}) are replaced by the corresponding hyperbolic functions, which correspond to the case (ii) of  $\tilde H$ in (\ref{specialcaseiii}). Similar results can be obtained for the case (iii).

When ${\kappa<0}$, the canonical representation of these spacetimes is ${\alpha = 1}$, ${\beta = 0}$, so that $ {\,P = 1+{\textstyle{1\over6}}\Lambda\zeta\bar\zeta \,}$, ${Q = 1-{\textstyle{1\over6}}\Lambda\zeta\bar\zeta \,}$ and ${\kappa = {\textstyle{1\over3}}\Lambda<0}$. In this case the transformation (\ref{transub1}), (\ref{transub2}) is simply
\begin{eqnarray}
  a &=& \frac{1}{\sqrt{2}}\,\frac{3}{|\Lambda|}\, \frac{1-\frac{1}{6}\Lambda\zeta\bar\zeta}{\sqrt{\zeta\bar\zeta}}\,, \\
  x &=&   \frac{1}{2}\, \sqrt{\frac{3}{|\Lambda|}}\,\,  \ln\left({\frac{\zeta}{\bar\zeta}}\right).  
\end{eqnarray}
Notice that $x$ is purely imaginary in this case. Indeed, introducing the polar parametrisation ${\zeta=\rho\,\exp(\im\varphi)}$,
the transformation becomes
\begin{equation}
a = \frac{1}{\sqrt{2}}\,\frac{3}{|\Lambda|}\, \frac{1-\frac{1}{6}\Lambda\rho^2}{\rho}\,, \qquad
x =  \im\, \sqrt{\frac{3}{|\Lambda|}}\,\,  \varphi.
\label{imagtransf}
\end{equation}
However, this is consistent with the form of the metric (\ref{Metric3d}). It follows from the relation (\ref{trsub1}) that the sign of the parameter $\kappa$ is the same as the sign of the function $k$, so that ${k<0}$ in this case, and the coordinate~$x$ must (formally) be purely imaginary. In fact, ${2k= \frac{1}{3}\Lambda\, a^2 + 1=\frac{1}{2}(3/\Lambda)((1+\frac{1}{6}\Lambda\rho^2)^2\rho^{-2}<0}$, and the expression (\ref{trsub2}) gives
\begin{equation}
\frac {1}{4k^2}\,{\d a}^2 + \frac{1}{2k}{\d x}^2 = \frac{2}{(1+\frac{1}{6}\Lambda\rho^2)^2}\,(\,\d\rho^2+\rho^2\d\varphi^2)\,.
\end{equation}
Again, the function $H$ for the case (ii) takes the form (\ref{KundtConflat}) in which
\begin{eqnarray}
 {\cal A}(u) &=&  \frac{6\sqrt2}{\Lambda} \Big( \gamma_2 (u) - \sqrt{{3}/{|\Lambda|}} \,\Big), \nonumber\\
 {\cal B}(u) &=&  \frac{-1}{\sqrt2} \,\gamma_1 (u) \left(\!\cosh\Big(\!\sqrt{{\textstyle{1\over6}}|\Lambda|}\,{\textstyle \int\!{\gamma_3}({u}) \,\d {u} } \Big)- \im \,\sinh\Big(\!\sqrt{{\textstyle{1\over6}}|\Lambda|}\,{\textstyle \int\!{\gamma_3}({u}) \,\d {u} } \Big) \!\right)\!, \nonumber\\
 {\cal C}(u) &=&  -{\textstyle \frac{1}{6}}\Lambda\, {\cal A}(u) - 2\sqrt{{6}/{|\Lambda|}}\,. \label{coeffiinegat}
\end{eqnarray}
Analogous results are also obtained for the case (iii) of (\ref{specialcaseiii}), provided a linear combination of the terms with different signs (that is with their upper and lower choices) is considered.

 \section{Summary and conclusions}
 
In this contribution we have analyzed pure radiation spacetimes which are of algebraic types~N and~O. In particular, we have concentrated on the metrics presented in recent works \cite{EdgarVickers:1999,EdgarRamos:2005,EdgarRamos:2007a,EdgarRamos:2007b}. We have found explicit transformations which put all these solutions to the Ozsv\'ath--Robinson--R\'ozga form (\ref{metricORR})--(\ref{Kundtk}), first given in~\cite{OzsvathRobinsonRozga:1985}. This conveniently represents an entire family of type~N or conformally flat Kundt spacetimes with an arbitrary cosmological constant~$\Lambda$, which are either vacuum or contain pure radiation.

By putting the above metrics into the unified coordinate system of (\ref{metricORR}) we have confirmed that their derivation based on the GIF formalism, as presented in \cite{EdgarVickers:1999,EdgarRamos:2005,EdgarRamos:2007a,EdgarRamos:2007b}, is mathematically correct. When compared with the general function (\ref{KundtConflat}) characterizing the complete family of conformally flat solutions, which involves four arbitrary (real) functions of $u$, we have identified only few apparent differences. In particular, in the solutions with ${\Lambda=0}$ given by (\ref{ozn0c}), instead of an arbitrary function ${{\cal C}(u)}$ there is a constant. Similarly, in the case ${\Lambda<0}$ such that ${\tau\bar{\tau}+\frac{1}{6}\Lambda=0}$, the function $B(u)$ in (\ref{ozn2c}) is  constant, namely ${B(u)=-1/(2\lambda^4)}$, but this can always be achieved by a coordinate transformation (see the footnote in section~\ref{sectypeonegat}). In the cases ${\Lambda\not=0}$ for which ${\tau\bar{\tau}+\frac{1}{6}\Lambda\not=0}$, the functions ${{\cal A}(u)}$ and ${{\cal C}(u)}$ in equations (\ref{Cverse2c}) and (\ref{coeffiinegat}) differ only by constants. Compared to a generic expression (\ref{KundtConflat}), this again does not represent any loss of generality. Indeed, coordinate freedom of the metric (\ref{metricORRspec}), namely \begin{equation}
v=v'f(u')+\frac{\dot f(u')}{\kappa}\,,\qquad u=\int\frac{\d u'}{f(u')}\,,
\end{equation}
implies 
\begin{equation}
{\cal A}'=\frac{\cal A}{f^2}-\alpha\,F\,,\quad  
  {\cal B}'=\frac{\cal B}{f^2}- \beta\,F\,,\quad
  {\cal C}'=\frac{\cal C}{f^2}+{\textstyle\frac{1}{6}}\Lambda\alpha\,F\,,
 \end{equation} 
where  ${\,F(u')=\frac{1}{\kappa}\left(\frac{\dot f^2}{f^2}-2\frac{\ddot f}{f}\right)
}$ and $f(u')$ is an arbitrary function. Consequently, ${{\cal C}'+{\textstyle\frac{1}{6}}{\cal A}'=f^{-2}({\cal C}+ {\textstyle\frac{1}{6}}\Lambda{\cal A})}$ in which $f$ can be prescribed arbitrarily. In addition, it follows from expression (\ref{PhiKundtO}) that the $u$-dependence of the pure radiation component~$\Phi_{22}$ is modified by such a coordinate transformation and can be set, e.g., to a constant.

However, it should be pointed out that the function $k$, introduced in equation (\ref{ozn3a}), may have \emph{negative} values when the cosmological constant is negative. In such a case, the metrics (\ref{Metric3a}) or (\ref{Metric3d}) do not maintain a correct signature, and roots of negative $k$ also occur in the original metric (\ref{Metric3a}). This problem disappears when the transformation (\ref{transub1}), (\ref{transub2}) to the Ozsv\'ath--Robinson--R\'ozga form (\ref{metricORRspec}) is performed (see also relations (\ref{imagtransf})).

Let us finally note that the classes of spacetimes studied in this contribution have very interesting geometrical properties. For example, they provide exceptional cases for the invariant classification of exact solutions, see e.g. \cite{Skea:1997,MilsonPelavas:2008}. It has also been demonstrated \cite{KoutrasMcIntosh:1996,BicakPravda:1998,Pravdaetal:2002} that, for conformally flat pure radiation (and some other type~N, III and~O) Kundt spacetimes, all scalar curvature invariants constructed from the Riemann tensor and its covariant derivatives of all orders identically vanish.

\begin{acknowledgements}
This work was supported by the grant GA\v{C}R~202/06/0041 and by the Czech Ministry of Education under the projects MSM0021610860 and LC06014.
\end{acknowledgements}

\end{document}